# Changing cell mechanics - a precondition for malignant transformation of oral squamous carcinoma cells

Felix Meinhövel[1,2], Roland Stange[4], Jörg Schnauß[4,5], Michael Sauer[6], Josef A. Käs[4] and Torsten W. Remmerbach[1,2,3]

[1] Section of Clinical & Experimental Oral Medicine, University Hospital, Leipzig University, Germany

[2] Department of Oral, Maxillofacial and Facial Plastic Surgery, University Hospital, Leipzig University, Germany

[3] Griffith Institute of Health, Griffith University, Queensland, Australia

[4] Faculty of Physics and Earth Sciences, Peter Debye Institute, Leipzig University, Germany

[5] Fraunhofer Institute for Cell Therapy and Immunology (IZI), DNA Nanodevices Group, Leipzig, Germany

[6] Department of Oral Maxillofacial and Facial Plastic Surgery, SRH Zentralklinikum Suhl GmbH, Suhl, Germany

Corresponding author: Prof. Dr. Josef A. Käs, Faculty of Physics and Earth Sciences, Peter Debye Institute, Leipzig University, Germany, Linnéstraße 5, D-04103 Leipzig, Germany; Email: jkaes@physik.uni-leipzig.de; Phone: +493419732470; Fax: +493419732479

**Abstract**

Oral squamous cell carcinomas (OSCC) are the 6$^{th}$ most common cancer and the diagnosis is often belated for a curative treatment. The reliable and early differentiation between healthy and diseased cells is the main aim of this study in order to improve the quality of the treatment and to understand tumour pathogenesis. Here, the optical stretcher is used to analyse mechanical properties of cells and their potential to serve as a marker for malignancy. Stretching experiments revealed for the first time that cells of primary OSCCs were deformed by 2.9 % rendering them softer than cells of healthy mucosa which were deformed only by 1.9 %. Furthermore, the relaxation behaviour of the cells revealed that these malignant cells exhibit a faster contraction than their benign counterparts. This suggests that deformability as well as relaxation behaviour can be used as distinct parameters to evaluate emerging differences between these benign and malignant cells. Since many studies in cancer research are performed with cancer cell lines rather than primary cells, we have compared the deformability and relaxation of both types, showing that long time culturing leads to softening of cells. The higher degree of deformability and relaxation behaviour can enable cancer cells to traverse tissue emphasizing that changes in cell architecture may be a potential precondition for malignant transformation. Respecting the fact that even short culture times have an essential effect on the significance of the results, the use of primary cells for further research is recommended. The distinction between malignant and benign cells would enable an early confirmation of cancer diagnoses by testing cell samples of suspect oral lesions.

**Abbreviations**

OSCC, oral squamous cell carcinoma; OPMD, oral potentially malignant disorders

**Introduction**

Oral squamous cell carcinomas (OSCCs) are the most common cancers of the head and neck. Usually, the OSCC develops on the basis of oral potentially malignant disorders (OPMD) [1–5]. In 50% - 60% of the cases the confirmation of diagnosis is belated, because the neoplasm has not been recognized or has been mistaken as harmless [6, 7]. Early diagnosis and treatment of this entity enables recovery. Consequently, an early and efficient clarification of doubtful lesions is requested to reduce the unacceptably high morbidity and mortality of oral squamous cell carcinomas [8]. Various in vitro studies based on cell lines of breast- and cervix carcinomas have shown that the viscoelastic properties of native and healthy cells compared to malignant cells differ significantly [9–11]. The filaments of the cytoskeleton (actin microfilaments, intermediate filaments and microtubules) form a structure that determines the shape and the mechanical properties of the cell [12–20]. In cancer cells, however, the neoplastic transformation is accompanied by a reduction of the cytoskeletal polymers, in particular actin-associated proteins that lead to a reduction of the structural integrity of the cell [21–24]. Results of a pilot study of our group indicate that this phenomenon seems to occur in well-established cell lines of the oral cavity. The work of Remmerbach et al. yielded first insights into biomechanical changes correlated to tumour growth, but at that time the optical stretcher technology did not allow to measure statistically relevant cell numbers [25]. The continuation of the studies revealed that the deformability of the cells is significantly influenced by culturing conditions and cell treatment before measurement [26]. In addition, it became increasingly clear that the cell deformability is temperature-dependent [27–29]. Consequently, we have used a well-defined culture protocol and an accurate temperature control during the measurements in order to acquire reproducible and statistically meaningful data.

This study considers the heterogeneity of the samples especially when originating from primary tissue. Tissue of the oral cavity constitutes a special case compared to most other tissues previously used for biomechanical measurements. It contains a large variety of cell types and a fraction of not fully differentiated cells since the tissue is self-renewing frequently. The measurements with the optical stretcher were performed in a way that cells were not pre-selected according to their morphology to mirror the heterogeneous cell populations also in the mechanical data.

An additional experiment investigated the comparability of cell lines derived from OSCCs to the primary samples. The cell lines are widely used in cancer research as a model system for cancer. With respect to the mechanical aspects there is a large doubt that those properties are preserved in cells that have undergone many passaging cycles on non-physiological substrates.

**Results**

To check the influence of cultivation on the primary cells, their stiffness was measured over a number of passage cycles. We found that cells with increasing passage number became increasingly softer (Figure 1A; significance tested with a t-test). After a few passages the cells stopped growing entirely. This effect could be determined for the primary samples of OSCC as well as for healthy mucosa. Since the growth duration of the samples was inconsistent, only the first six passages are shown (Figure 1A). Furthermore, the proliferation rate of the cells was studied by measuring their doubling time. Figure 1B shows that the proliferation rate increases with increasing passage number, before the growth was stalled globally since cells lost the ability to attach to the surface (and grow) possibly due to the repetitive treatment with trypsin. This phenomenon appeared almost consistently for all samples.

The expectation that cancer cells proliferate faster than healthy oral mucosa cells could not be confirmed in this study [35–37]. Doubling times of healthy oral mucosa cells are comparable or even shorter than the doubling time of oral squamous cell carcinoma. The period from initial cultivating to first passage has been excluded from analysis since the cells from different pieces of tissue require a varying time to adhere – a phenomenon that is not understood yet. A defined amount of single cells was seeded after the first passage in order to provide comparable conditions. In our experiments, cells were grown to 60 - 70 % confluence before passaging, which was reached with the healthy control cells after seven to eighteen days whereas the cancer cells needed five to twenty days (Figure 1B).

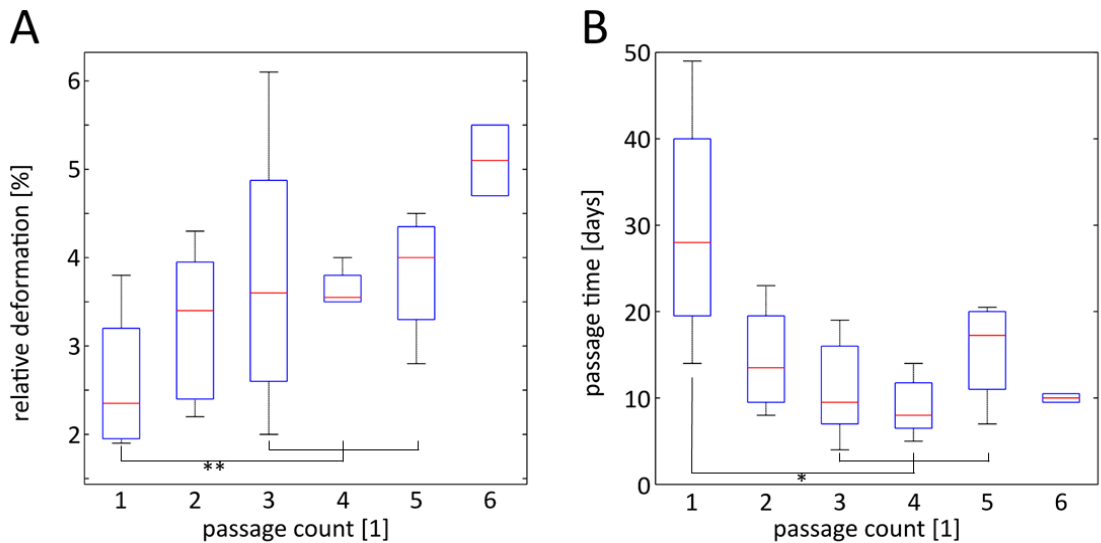

**Figure 1.** Overview of all measurements of primary samples. (A) The boxplot shows the relative deformation at the end of the stretching phase with regard to the passage count, with the whiskers representing the error bars. With increasing passage number, the primary cells become softer. (B) Plotting the passage time vs. passage number reveals a increasing proliferation rate with increasing passage number.

In contrast to the cells from commonly used cell lines (CAL 27, CAL 33, BHY), the primary cells (benign as well as cancerous) are very heterogeneous due to their varying sizes, different stages in the cell cycle and occasionally appearing microtentacles (Figure 2A). This already becomes apparent when comparing the distribution of the cell sizes. While healthy and cancerous primary cells have a comparable size, cells from cell lines are in average 2 µm smaller and sizes spread less compared to primary samples (Figure 2A). This is especially important since the cell size directly affects the deformability showing that larger cells are more deformed than smaller cells (Figure 2A). The different average cell size as well as the larger spread of sizes in primary tissue directly reflects the variety of cells from a heterogeneous tissue like the oral cavity and even more the OSCC. Only outliers with obvious measurement artefacts were excluded from data analysis of the deformability measurements to avoid bias and preselection.

The differences in deformability are represented in Figure 2B as the relative change of ellipticity during optical stretcher measurements. On average, the healthy samples (n= 360) have a relative change in ellipticity and deformation of about 1.9 % compared to 2.9 % in primary cancer cells (n= 1465) at the end of the stretch. The statistical evaluation revealed a highly significant difference between both tissue types (p < 0.001) displaying that cell originating from cancerous tissue are softer than their healthy counterparts. Since their sizes are comparable, the effect is not based only on geometrical arguments.

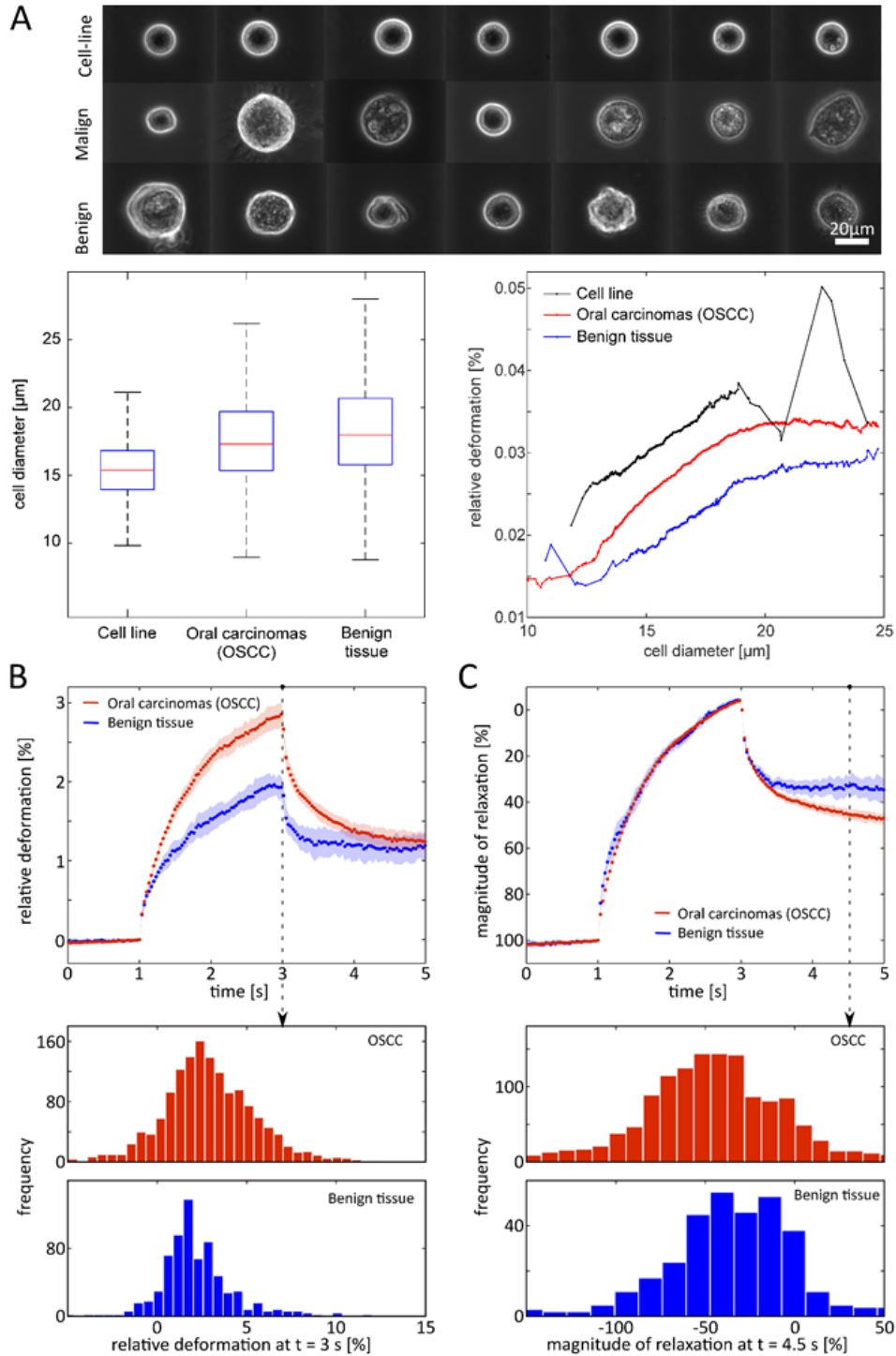

**Figure 2.** (A) Representative examples for the diversity of cells. The cells in the upper line originate from cancer cell lines (CAL 27, CAL 33, BHY). The cells below originate from primary cancerous tissue (middle) and primary benign tissue (bottom). The boxplot illustrates that cell originating from primary tissue are slightly larger and show a broader size distribution compared to cell lines. Cells from primary tissue as well as from cell lines display that larger cells are more deformed than smaller ones. (B) Comparing the relative deformability of primary OSCC (red; Passage 0; n= 1465) and benign oral epithelial tissue (blue; P0; n= 360) reveals that malignant cells are significantly softer than their benign counterparts. At the end of the stretching (dashed line; t = 3 s) the counts of the relative deformations show the distribution of the maximum deformability within the distinct cell population in the according histograms. (C) The magnitude of relaxation for primary OSCC (red) and primary oral epithelial tissue (blue), derived from the relative change in ellipticity normalized to the end of stretching (t = 3 s), illustrates that malignant OSCC cells were still contracting after

2 s while malignant cells seem to reach a stable plateau. Representative data after a total observation time of t = 4.5 s are shown in the according histograms displaying the distribution of relaxation values.

However, the maximum deformation at the end of the deformation period is not the only meaningful parameter which can be extracted from the measurements since the Optical Stretcher provides a very good time resolution of the whole deformation process. Evaluating the relaxation behaviour after deformation allows to study an important cytoskeletal property of cells - their viscoelastic behaviour triggered by external stresses – showing that cells do not return to their original shape within the observation time. Investigating the relaxation of cells after the stretching procedure revealed further differences between malignant and benign tissue. Tumour cells do not only become softer (Figure 2B) but also return faster towards their original shape than healthy cells (Figure 2B & C).

When normalizing the curves to the maximal deformation (Figure 2C), it becomes apparent that cancerous and healthy cells initially relax comparably for the first 0.4 s after the stretching phase. Subsequently, the relaxation of healthy cells is almost completely halted and plateaus at still high deformability values (compared to the maximal deformation) while cancer cells continued to relax towards their initial shape over the entire relaxation period (t = 3-5 s). Since the behaviours are comparable in the initial relaxation phase but diverge later, it can be expected that the difference is not a pure requisite of the magnitude of the deformation but rather a result of differing internal dynamics. Generally, the statistical evaluation revealed a highly significant difference between healthy probes and primary cancer samples (p < 0.001) implying that primary cancer cells differ not only in their maximum deformability from healthy probes but also in their relaxation behaviour.

Deformability curves of the OSCC cell lines (CAL 27, CAL 33, BHY) are presented in Figure 3. They all exhibit a higher deformability than the primary samples, which at first sight appears counterintuitive since cells from cell lines are smaller than the primary cells (Figure 2A) and should deform less. However, they are significantly softer, which is in accordance with the observed effect that primary samples become softer with increasing passage number (Figure 1). In contrast, the relaxation behaviour is decreasing compared to the primary samples.

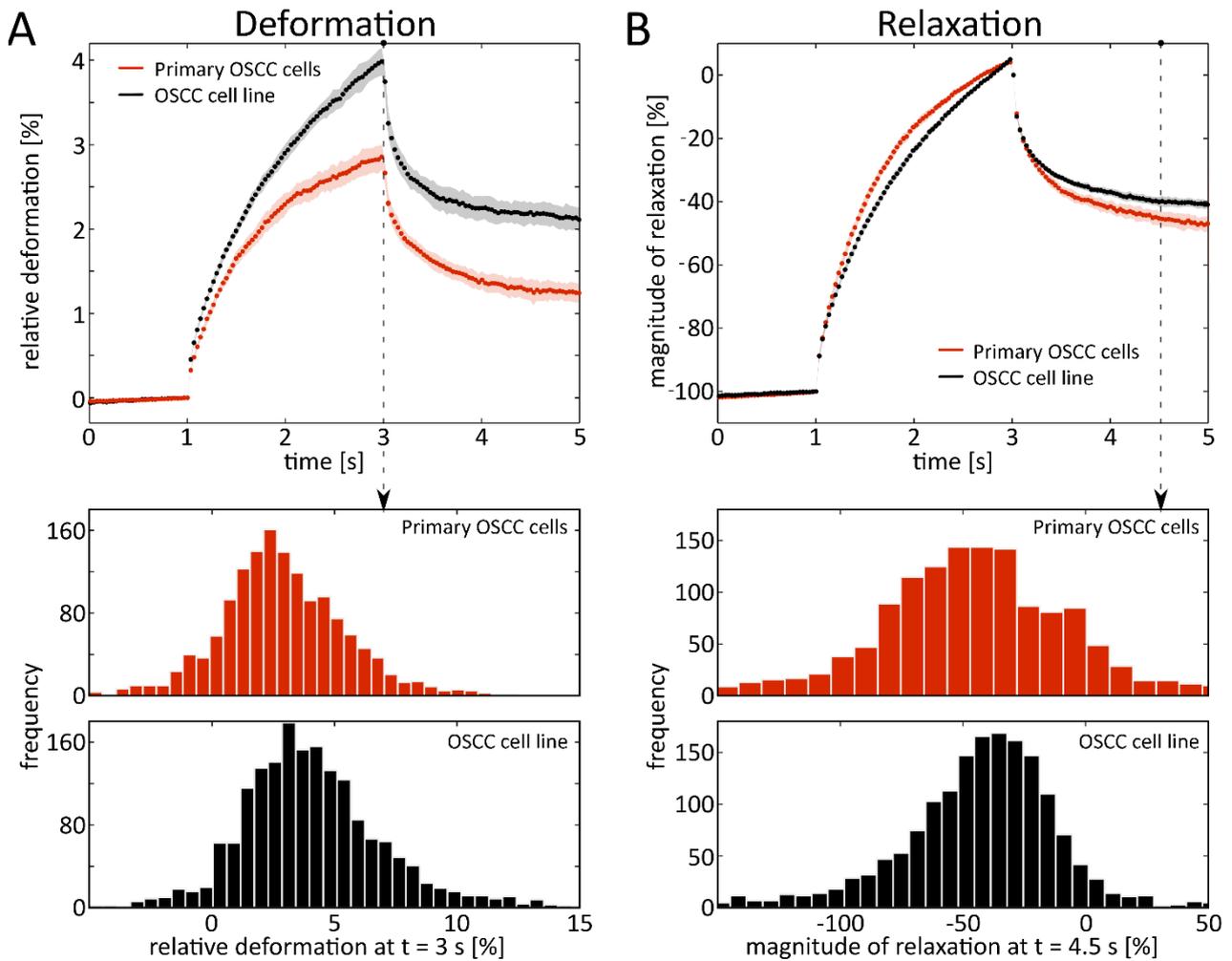

**Figure 3.** (A) A comparison of the relative deformability of primary OSCC cells (red; Passage 0; n= 1465) and OSCC cell lines (back; pooled, n= 1780) reveals a much higher deformability for cells from OSCC cell lines. Histograms show the distribution of the maximum deformability (dashed line; end of stretching at t = 3 s) within the cell population and (B) the magnitude of relaxation (dashed line; t = 4.5 s) for primary OSCC (red) and OSCC cell lines (black). Cell lines exhibit a smaller relaxation than primary OSCC cells, which is, however, still larger than for the relaxation of primary benign cells. Histograms show the distribution of relaxation for both populations at t = 4.5 s.

**Discussion**

This is the first representative study that investigates the biomechanical properties of primary cells of the oral cavity measured with the optical stretcher without preselecting any homogeneous subpopulations of cells. Even after some time in culture, the primary cells exhibit a higher heterogeneity than the commonly used cell lines, which suggests that some of the differentiated properties of the tissue cells are preserved.

The optical stretcher was used as a diagnostic tool to analyse mechanical properties of the cytoskeleton of single cells from primary OSCCs and healthy control tissue. Results of both tissue sample types were compared to measurements of the standard cell lines being used as model systems for OSCCs. The results clearly indicate that healthy oral mucosa cells exhibit significantly different mechanical properties compared to malignant cells. Benign cells are stiffer and therefore less deformable than cancerous primary tissue cells. In addition, it has been shown that the relaxation of the cells after active deformation is a second parameter which can be used to analyse mechanical differences: The evaluation of these data revealed that tumour cells have a greater ability to return to their initial state than healthy cells, which have a noticeable plastic deformation behaviour. Thus, while the deformation experiments show that tumour cells are softer with more pronounced viscous contributions and appear more fluid-like than healthy cell, they appear more elastic than healthy cells during the relaxation phase. They display a continuous return towards their initial shape - a characteristic of an elastic deformation – while healthy cells quickly plateau at still high deformability values. Consequently, the question arises what kind of mechanisms are responsible for the different mechanical behaviour of cells in the tumour cluster and why these differences are essential for malignancy? It can be speculated that plasticity effects such as bond breaking during the deformation or active responses superimpose classical viscoelastic effects. Due to the multitude of variables in these systems, a comprehensive modelling of arising behaviours is not auxiliary to gain new insights, which unfortunately impedes a direct translation of optical deformability into the more conventional metrics of moduli.

From the proliferation assays presented in this study, it can be excluded that malignant cells are softer solely due to a higher proliferation rate and therefore to a lack of time for cytoskeleton formation. Our results clearly show that the proliferation rates of primary benign and malignant cells do not significantly differ in vitro.

Currently, there is a controversial discussion whether tumour cells are already soft in the tumour or if softening can be interpreted as a progressive course of disease leading to metastatic tumours [38]. Due to the fact that the cells in our experiments were removed from solid primary tumours and not from metastases or already metastasised tumours (all tumours G2 (moderately differentiated) and M0 (no distant metastasis); see supplemental table 1), it is obvious that the primary tumour cells in tumour tissue already show changed mechanical properties such as deformability and relaxation.

The property of stronger contractility combined with a more elastic behaviour could be essential for the malignancy of cells that have the potential to metastasise or to traverse tissue as a pathogenic factor. Therefore, we speculate that the different mechanical behaviour may be a result of the transformation from benign to malignant tissue. It can be also speculated that the softening of the cells is a necessary alteration or a precondition for tumour aggressiveness that may facilitate the traverse in cell clusters to be reshaped as soon as possible.

The difference of mechanical properties concerning cell lines seems to have another origin. We have observed that primary cells show decreasing cell stiffness with increasing passage number. This trend matches with the findings from experiments with established cell lines, where a greater deformability and accordingly softer cells were detected. Considering that most cell lines, including the cell lines used in this study, were subject to numerous passages and are in cell culture for many years, comparable data between primary tissue cells and established cell lines cannot be expected. The effect of softer behaviour with increasing passage number might be induced by the unphysiological micro-environment of in vitro culture conditions. These cells do not grow in their natural environment with mediators etc., but on a rigid plastic surface with an artificial medium and repetitive enzymatic treatment. Thus, to establish a serious diagnostic tool for direct application in the medical practise, the use of primary samples without cultivation is strongly recommended. We would like to note that cell lines are in fact robust test systems for new diagnostic and/or therapeutic platforms, especially since the cells are more homogenous and comparable than primary cells. However, for further investigations of tumour related cell softening, we suggest to measure the malignant tissue as well as a benign tissue sample in pairs originating from the same patient and omit direct comparisons of primary malignant cells with benign cell lines.

In conclusion, we were able to show that healthy and malignant oral tissue can be significantly distinguished from one another. They can be discriminated by their deformation as well as their relaxation behaviour. Our results indicate that the change of the cytoskeletal properties is rather a precondition for malignant transformation than the outcome of the transformation from the healthy to the diseased cell. These findings strongly emphasize the need to study if the deformability can be applied as an area-wide diagnostic tool for an early diagnosis of OSCCs.

**Materials and methods**

*Acquisition of primary cells*

The ethics committee of the University of Leipzig has reviewed and approved the research protocol (Reg.-No. 193/2003) and all participants gave written informed consent according to the Declaration of Helsinki. The tissue samples of the OSCC were generated during tumour resection. A small part of the resected tumour was separated and put into a centrifuge tube, containing 5 ml of Keratinocyte Basal Medium (Clonetics, Cologne, Germany) with 100 units/ml penicillin, 100 µg/ml streptomycin and 10 µg/ml amphotericin. The specimens of healthy donors were acquired during the surgical removal of third molars and were treated in the same way as the tumour specimens.

*Cell lines*

The established and commercially available cell lines CAL 27, CAL 33 [30] and BHY [31] were used for this study. All cell lines were obtained directly from the cell bank DSMZ (Deutsche Sammlung von Mikroorganismen und Zellkulturen GmbH, Braunschweig, Germany), who performed cell line characterizations. All cell lines were passaged in our laboratory for less than 6 months after resuscitation.

*Cultivation protocol*

The tissue samples were washed in Betaisodona solution (Mundipharma, Limburg/ Lahn, Germany) for five minutes for disinfection, rinsed in phosphate buffered saline, and cut into pieces of about 1 mm³. These pieces were placed in an empty cell culture flask. After about 30 minutes, 20 ml of Keratinocyte Basal Medium (Clonetics, Cologne, Germany) with 100 units/ml penicillin, 100 µg/ml streptomycin and 10 µg/ml amphotericin were added and incubated at 37 °C/ 5 % $CO_2$. The primary cultivation time varied from 14 to 49 days until 60 - 70 % confluence was reached. Prior to the experiment or to passaging the cells were detached by incubation with 2 ml of 0.25 % trypsin/EDTA for five minutes at 37 °C/ 5 % $CO_2$. The trypsin reaction was stopped by adding 2 ml of DMEM (10% serum) to the cell suspension. To remove cell conglomerates and dead cells the suspension was strained through a 40 µm cell strainer (Falcon™ Cell

Strainers No.: 352340) and centrifuged twice for 2 min at RCF = 200 g at room temperature. Subsequently, the pellet was resuspended in 1 ml Keratinocyte Basal Medium (Clonetics, Cologne, Germany) with 100 units/ml penicillin, 100 µg/ml streptomycin and 10 µg/ml amphotericin.

*Optical stretcher*

It has been shown in several experiments (carcinoma of cervix or breast [11]), that the optical stretcher produces relevant data to differentiate between cancerous and healthy tissue.

With the two-beam optical stretcher single cells in suspension were initially trapped with a laser power of 100 mW using two facing and slightly divergent laser beams [9, 11, 20, 23, 32, 33]. The cell suspension was given into a microfluidic system to channel a single cell into the optical trap. Applying higher laser power of 800 mW for two seconds, the momentum transfer at the interface leads to visible deformation of the cell along the laser axis. After deformation of the cell, the ability of retraction was analysed for two seconds [9]. A series of images was recorded during each measurement for further analyses of cell deformation and retraction. Therefore, the images were analysed automatically with a custom-made image analysis software (MATLAB 7.11.1, The Mathworks, Natick, Massachusetts, USA) using edge detection processes to determine the cell shape. To quantify the deformation, the ellipticity of the cells was extracted by fitting an ellipse on the edge points. The relative change in ellipticity (relative deformability) allows to compare different measurements (e.g. tumour1 vs. benign control).

*Statistics*

Biological cells are well known for their adaptability and manifoldness that do not necessarily arise randomly. Thus the frequency distribution of the deformability is not Gaussian. Consequently, median values were calculated and plotted for statistical evaluation. To generate confidence intervals for the non-parametric distributions we used the bootstrapping method; all confidence intervals indicate the 95% range. The Wilcoxon Rank sum test was used to check whether the values originate from the same basic distribution [34] as a measure for statistical significance.

**Acknowledgement:** We would like to thank the pathologist Dr. Schütze for the histopathological classification of all cancer samples. We also thank our technician Mrs Tröger for her support concerning cell culture and Ms Milani for her critical remarks. We acknowledge the financial supported by an R&D grant (SAB, Project 9889/1519) from the European Fund for Regional Development (EFRE) 2000-2006 and the state of Saxony, as well as the European Research Council (ERC-741350).

**Compliance with Ethical Standards**

**Funding:** We acknowledge the financial supported by an R&D grant (SAB, Project 9889/1519) from the European Fund for Regional Development (EFRE) 2000-2006 and the state of Saxony, as well as the European Research Council (ERC-741350).

**Potential conflict of interest:** One of the authors (JAK) holds a patent on the optical stretcher technique and consults on its potential applications. One author (TWR) is CEO of the DGOD Deutsche Gesellschaft für orale Diagnostika mbH, Leipzig, Germany (Dental healthcare supplier). The other authors declare no competing financial interests.

**Ethical approval:** All procedures performed in studies involving human participants were in accordance with the ethical standards of the institutional and/or national research committee and with the 1964 Helsinki declaration and its later amendments or comparable ethical standards. (Reg.-No. 193/2003)

**Informed consent:** Informed consent was obtained from all individual participants included in the study.